\documentclass[preprint,aps,12pt,showpacs,nofootinbib,tightenlines]{revtex4}
\usepackage{mathrsfs}
\usepackage{amsmath}
\usepackage{amssymb}
\usepackage{epsfig}
\usepackage{graphicx}
\newcommand{\psl}{ P \hspace{-2.4truemm}/ }
\newcommand{\nsl}{ v \hspace{-2.2truemm}/ }

\newcommand{\ksl}{ k \hspace{-2.2truemm}/ }
\newcommand{\lsl}{ l \hspace{-2.2truemm}/ }

\def\be{\begin{eqnarray}}
\def\en{\end{eqnarray}}
\def\non{\nonumber\\}

\begin{document}
\title{Next-to-Leading-logarithm threshold resummation for exclusive $B$ meson decays}
\author{Zhi-Qing Zhang
\footnote{zhangzhiqing@haut.edu.cn}} 
\affiliation{Department of Physics, Henan University of Technology,
Zhengzhou, Henan 450052, P. R. China}
\author{Hsiang-nan Li
\footnote{hnli@phys.sinica.edu.tw}} 
\affiliation{Institute of Physics, Academia Sinica,
Taipei, Taiwan 115, Republic of China} 
\date{\today}
\vskip 2cm

\begin{abstract}
We extend the threshold resummation of the large logarithms $\ln x$ which
appear in factorization formulas for exclusive $B$ meson decays, $x$ being a
spectator momentum fraction, to the next-to-leading-logarithm (NLL) accuracy.
It is shown that the NLL resummation effect provides suppression in the end-point
region with $x\sim 0$ stronger than the leading-logarithm (LL) one,
and thus improves perturbative analyses of the above processes.
We revisit the $B\to K\pi$ decays under
the NLL resummation, and find that it induces 20-25\% variation of the
direct CP asymmetries compared to those from the LL resummation. Our
way to avoid the Landau singularity in the inverse Mellin transformation
causes little theoretical uncertainty.
\end{abstract}

\vspace{1cm}

\maketitle

\section{Introduction}

Factorization theorems have been one of the major theoretical approaches to exclusive
$B$ meson decays, in which a decay process is factorized into a convolution of a
hard kernel with hadron distribution amplitudes. A crucial issue on the application
of factorization theorems to a key ingredient of these decays, $B$ meson transition
form factors, is the end-point singularity, which appears as a spectator parton carries
a vanishing momentum fraction $x\to 0$ \cite{SHB,ASY,BF}. Because of this end-point
singularity, a $B$ meson transition form factor is treated as a nonperturbative input
in the QCD-improved factorization approach \cite{BBNS} based on the collinear
factorization theorem. In the soft-collinear effective
theory the end-point singularity can be removed by the zero-bin subtraction
\cite{Manohar:2006nz}, so that a $B$ meson transition form factor becomes factorizable
in the collinear factorization. It was argued that a parton transverse momentum $k_T$
is not negligible, when the end-point region is important. The perturbative QCD approach
based on the $k_T$ factorization theorem was then proposed \cite{LY1,KLS,LUY}, in which the
end-point singularity is regularized by a parton $k_T$, and a $B$ meson transition form
factor also becomes factorizable \cite{TLS}.

An alternative removal of the end-point singularity in the framework of the collinear
factorization has been suggested in \cite{Li:2001ay}. When the end-point region dominates,
the double logarithms $\alpha_s\ln^2 x$ from radiative corrections
\cite{ASY,KPY} should be organized to all orders
to improve perturbative expansion. The first systematic study was done in
\cite{Li:2001ay}, where these double logarithms were factorized from
exclusive $B$ meson decays
into a universal jet function, and resummed up to the leading-logarithm (LL)
accuracy. It was then shown that the resultant jet function
vanishes quickly at $x\to 0$, and suppresses the end-point singularities in the $B\to\pi$
form factors. The threshold resummation effect on more complicated two-body
hadronic $B$ meson decays, which involve the annihilation and nonfactorizable
amplitudes in addition to the factorizable one proportional to a transition
form factor, was analysed in \cite{Li:2002mi} and implemented in the PQCD approach
widely afterwards.

In this paper we will extend the LL threshold resummation performed in \cite{Li:2001ay}
to the next-to-leading-logarithm (NLL) accuracy. To accomplish this task, we calculate the
jet function stated above at one loop to identify the complete large logarithms, solve
an evolution equation for the jet function in the Mellin space to get all-order
summation of the logarithms, match the all-order summation to the one-loop result
to determine the initial condition of the jet function, and follow the best fit method in \cite{TLS}
to obtain the threshold resummaiton in the momentum fraction $x$ space. It will be demonstrated
that the NLL jet function exhibits suppression at the end point $x\sim 0$
stronger than the LL one. Because the threshold resummation modifies hard decay
kernels by including partial higher order contributions, hadron distribution amplitudes, such as
the uncertain $B$ meson distribution amplitude, should be
adjusted accordingly to maintain $B$ meson transition form factors \cite{Li:2005kt}.
Therefore, we compare the LL and NLL resummation effects by investigating their impacts on the
CP asymmetries in the $B\to K\pi$ decays, which are less sensitive to choices
of hadron distribution amplitudes. It is found that the replacement of the LL jet function by the
NLL one in the PQCD factorization formulas causes about 20-25\% variation of
the $B\to K\pi$ direct CP asymmetries, an effect which needs to be taken into account
for precision calculations of $B$ meson decays.

In Sec.~II we compute the one-loop jet function, derive its evolution equation
and the corresponding evolution kernels in the Mellin space, and solve the
evolution equation to attain the NLL threshold resummation.
The best fit method is employed to transform the jet function from the
Mellin space back to the momentum fraction space in Sec.~III. It is verified that
the Landau singularity in the inverse Mellin transformation
can be avoided, and the theoretical uncertainty is under control in the above method.
The impacts of the LL and NLL resummations on the CP
asymmetries in the $B\to K\pi$ decays are then
examined and compared. Section~IV contains the conclusion and outlook.

\section{THRESHOLD RESUMMATION}

\subsection{One-loop Quark Diagrams}
\begin{figure}[!]
\begin{center}
\vspace{-3cm} \hspace{1cm}\includegraphics[scale=0.8]{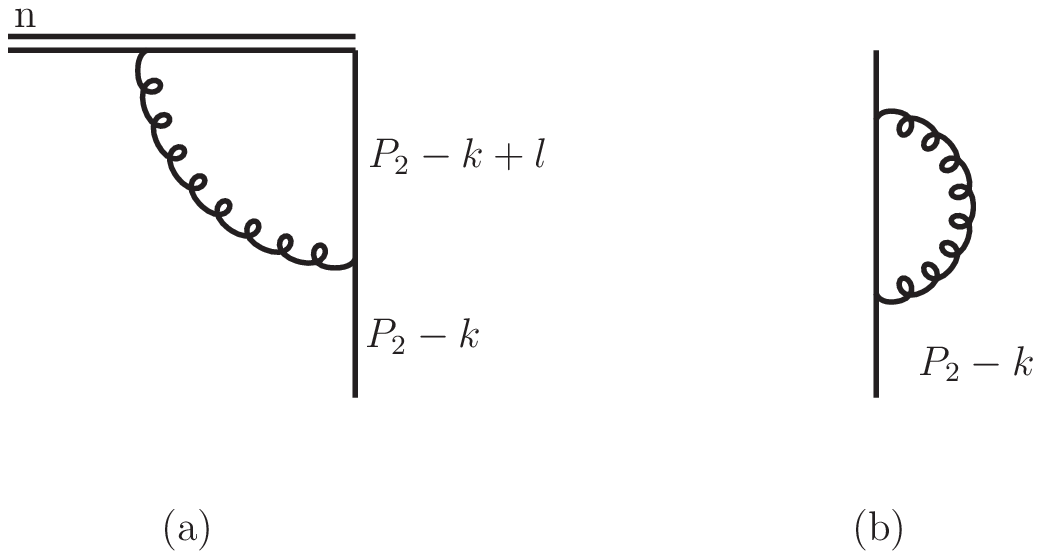}
\vspace{-17cm} \caption{One-loop diagrams for the jet function,
where the double line represents the Wilson link.}\label{full}
\end{center}
\end{figure}

The definition of the jet function $J(x)$ in terms of a quark field and its associated
Wilson link, which is constructed from the factorization of the radiative decay
$B(P_1)\to\gamma(P_2) l\bar\nu$, is referred to \cite{Li:2001ay}. The Wilson link
runs in the direction $n$, that contains the arbitrary components $n^+$ and $n^-$.
The quark momentum has been parametrized as $P_2-k$, where the photon momentum $P_2$ is in
the minus direction and the momentum $k=(xP_1^+,0,{\bf 0}_T)$ of the light quark
in the $B$ meson is in the plus direction. That is, this quark is slightly
off-shell by $(P_2-k)^2=-2x P_1^+P_2^-\equiv -xQ^2$. The leading-order (LO) jet function
has been chosen as $J^{(0)}(x)=1$. The one-loop vertex correction in Fig.~\ref{full}
is written as
\be
J^{(1)}_a(x)&=&-ig^2C_F\mu^{2\epsilon}\int\frac{d^{4-2\epsilon}l}{(2\pi)^{4-2\epsilon}}tr\left[\frac{\nsl_+\nsl_-}{4}\gamma_{\beta}\frac{\psl_2-\ksl+\lsl}{(P_2-k+l)^2}\right]\frac{n^\beta}{n\cdot ll^2},
\en
where $C_F=4/3$ is a color factor, and $\mu$ is the renormalization
scale. The projector $\nsl_+\nsl_-/4$, with the light-like vectors $v_+=(1,0,{\bf 0}_T)$
and $v_-=(0,1,{\bf 0}_T)$, arises from the
factorization of the jet function \cite{Li:2001ay}. A straightforward evaluation gives
\be
J^{(1)}_a(x)&=&-\frac{\alpha_sC_F}{4\pi}\left(-\frac{1}{\epsilon}-\ln\frac{4\pi\mu^2}{xQ^2}
+\ln^2\frac{Q^2x}{\xi^2}+\ln\frac{Q^2x}{\xi^2}
+\frac{4}{3}\pi^2+\gamma_E-2\right),\label{ma1}
\en
for $n^2>0$, with the $n$-dependent factor $\xi^2\equiv 4(P_2\cdot n)^2/n^2$
and the Euler constant $\gamma_E$. Note that the jet function depends on the Lorentz invariants
$(P_2-k)\cdot n\approx P_2\cdot n$ in the small $x$ limit and $n^2$, and that the Feynman rules
associated with the Wilson link shows a scale invariance in $n$. These facts
explain why the two vectors $P_2-k$ and $n$ appear via the ratio $\xi^2$ in Eq.~(\ref{ma1}).

The self-energy correction in Fig.~\ref{full} is expressed as
\be
J^{(1)}_b(x)
&=&-ig^2C_F\mu^{2\epsilon}\int\frac{d^{4-2\epsilon}l}{(2\pi)^{4-2\epsilon}}tr\left[\frac{\nsl_+\nsl_-}{4}
\gamma^\nu\frac{\psl_2-\ksl-\lsl}{(P_2-k-l)^2}\gamma_\nu\frac{\psl_2-\ksl}{(P_2-k)^2}\right]\frac{1}{l^2}\non
&=&-\frac{\alpha_sC_F}{4\pi}\left(\frac{1}{\epsilon}+\ln\frac{4\pi\mu^2}{xQ^2}
-\gamma_E+2\right).
\en
We then have the $\mathcal{O}(\alpha_s)$ jet function
\be
J^{(1)}(x)=J^{(1)}_a(x)+J^{(1)}_b(x)=-\frac{\alpha_sC_F}{4\pi}\left(\ln^2\frac{Q^2x}{\xi^2}
+\ln\frac{Q^2x}{\xi^2}+\frac{4}{3}\pi^2\right),\label{j1}
\en
which is independent of $\mu$, ie., ultraviolet finite.

We apply the Mellin transformation
from the momentum fraction $x$ space to the moment $N$ space
\be
\tilde{J}(N)\equiv\int^1_0dx(1-x)^{N-1}J(x). \label{mellin}
\en
It implies that the transformed jet function $\tilde{J}(N)$ at large $N$ collects the contribution
mainly from the small $x$ region.
The Mellin transformation of the LO jet function, $\tilde{J}^{(0)}(N)=1/N$, is trivial.
The Mellin transformation of Eq.~(\ref{j1}) yields, in terms of the variable
$\bar N\equiv N\exp(\gamma_E)$,
\begin{eqnarray}
\tilde{J}^{(1)}(N)&\approx&
-\frac{\alpha_sC_F}{4\pi}\left(\ln^2 \frac{Q^2}{\xi^2\bar N}+
\ln \frac{Q^2}{\xi^2\bar N}+\frac{3}{2}\pi^2\right)\frac{1}{N},\label{j2}
\end{eqnarray}
in the large $N$ limit up to corrections down by powers of $1/N$.

\subsection{Evolution Equation for $J$}\label{eff}

As indicated by the above one-loop calculation, the important logarithms in the jet
function depend on the factor $\xi^2\equiv 4\nu^2 P_2^{-2}$. To resum these logarithms,
we construct the evolution equation for the jet function \cite{Li:1996gi,Li:2013ela}
\begin{eqnarray}
2\nu^2\frac{dJ}{d\nu^2}=-\frac{n^2}{v_-\cdot n}v_{-\alpha}\frac{dJ}{dn^\alpha}.
\end{eqnarray}
The derivative respect to $n^\alpha$ applies to the Feynman rules of the Wilson
link, generating
\begin{eqnarray}
-\frac{n^2}{v_-\cdot n}v_{-\alpha}\frac{d}{dn^\alpha}\frac{n^\beta}{n\cdot l}
=\frac{\hat{n}^\beta}{n\cdot l},
\end{eqnarray}
with the special vertex
\begin{eqnarray}
\hat{n}^\beta=\frac{n^2}{v_-\cdot n}\left(\frac{v_-\cdot l}{n\cdot l}n^\beta-v_-^\beta\right).
\end{eqnarray}
The technique of varying Wilson links has been applied to the resummation of various
types of logarithms, such as the rapidity logaritms in the $B$ meson wave function \cite{Li:2012md},
and the joint logarithms in the pion wave function \cite{Li:2013xna}.
The Ward identity for the summation over the special vertices leads to the factorization
of the soft function $K$ and the hard function $G$ from the derivative of the jet function
\cite{Li:1996gi,Li:2013ela}
\begin{eqnarray}
2\nu^2\frac{dJ}{d\nu^2}=(K+G)\otimes J.\label{evo}
\end{eqnarray}

\begin{figure}[!]
\begin{center}
\includegraphics[scale=0.6]{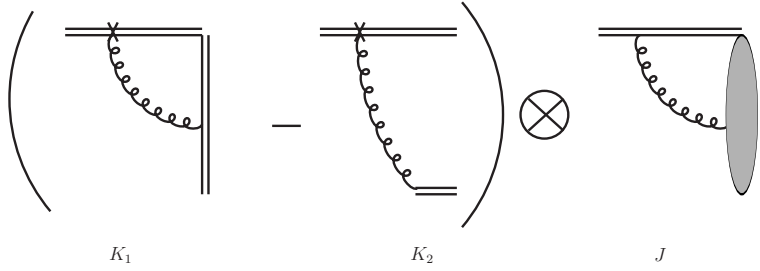}
\vspace{-13cm} \caption{Convolution of the $\mathcal{O}(\alpha_s)$
soft function $K$ with the jet function $J$, where the symbol $\times$
represents the special vertex.}\label{soft}
\end{center}
\end{figure}

Figure~\ref{soft} depicts the factorization of the soft function $K$ at $\mathcal{O}(\alpha_s)$,
which contains two pieces $K_1$ and $K_2$. The former is written as
\be
K_1&=&-ig^2C_F\mu^{2\epsilon}\int \frac{d^{4-2\epsilon}}{(2\pi)^{4-2\epsilon}}\frac{\hat{n}_{\mu}}{n\cdot l}\frac{g^{\mu\nu}}{l^2-m^2}\frac{P_{J\nu}}{P_{J}\cdot l}-\delta K,
\en
with the momentum $P_J=P_2-k$, where the gluon mass $m^2$, serving as an infrared regulator,
will approach to zero eventually. Choosing the additive counterterm
\be
\delta K=-\frac{\alpha_sC_F}{2\pi}\left[\ln (4\pi\nu^2)+\frac{1}{\epsilon}-\gamma_E\right],
\en
we have
\be
K_1&=&-\frac{\alpha_sC_F}{2\pi}\ln\frac{\mu^2}{\nu^2m^2}.\label{K1}
\en

The loop momentum $l$ flows through the jet function for $K_2$, so they
appear in a convolution
\be
K_2\otimes J&=&ig^2C_F\int \frac{d^4l}{(2\pi)^4}\frac{\hat{n}_\mu}{n\cdot l}
\frac{g^{\mu\nu}}{l^2-m^2}\frac{P_{J\nu}}{P_J\cdot l}J(x-\frac{l^+}{P^+_1}).
\en
Performing the contour integration over $l^-$ for $-(1-x)P^+_1<l^+<0$ followed by
the integration over the transverse momentum $l_T$, and employing the variable
change $u=x-l^+/P^+_1$, we arrive at
\be
K_2\otimes J
&=&\frac{\alpha_sC_F}{\pi}\left[\int^1_x du\frac{J(u)-J(x)}{u-x}
+\int^1_x du\frac{(u-x)J(x)}{(u-x)^2+M^2}\right],
\label{k2j}
\en
with the infrared regulator $M^2\equiv \nu^2 m^2/P^{+2}_1$.

The Mellin transformation of the first integral in Eq.~(\ref{k2j}) gives
\be
&&\int^1_0dx(1-x)^{N-1}\int^1_xdu\frac{J(u)-J(x)}{u-x}\non
&\approx&\int^1_0du J(u)\int^1_0dt\frac{(1-t)^{N-1}}{t}\left(\frac{1}{1-t}-u\right)^{N-1}
-\int^1_0dx(1-x)^{N-1}J(x)\int^{1}_0\frac{dt}{t},\label{t1}
\en
where the order of the integrations over $x$ and $u$ has been exchanged in the first term,
the variable change $x=u(1-t)$ has been applied, and the upper bound $1-x$ of the
integration variable $t$ in the second term has been approximated by 1. This approximation holds
up to an infrared finite constant, which will be compensated by matching later.
The further approximation $[1/(1-t)-u]^{N-1}\approx (1-u)^{N-1}$,
which holds in the dominant small $t$ region, brings Eq.~(\ref{t1}) into
\be
\tilde{J}(N)\int^1_0dt\frac{(1-t)^{N-1}-1}{t}
&\approx& -\tilde{J}(N)\ln \bar N.\label{k21}
\en
The above result is subject to corrections down by a power of $1/N$.

We also apply the Mellin transformation
to the second integral in Eq.~(\ref{k2j}):
\be
\int^1_0dx(1-x)^{N-1}J(x)\int^1_0du\frac{(u-x)}{(u-x)^2+M^2}
\approx\tilde{J}(N)\ln\frac{1}{M}. \label{k22}
\en
The sum of Eqs.~(\ref{k21}) and (\ref{k22}) yields the Mellin transformation of Eq.~(\ref{k2j}),
\be
\int^1_0dx(1-x)^{N-1} K_2\otimes J&=&
-\frac{\alpha_sC_F}{\pi}\left(\ln \bar N-\ln\frac{P^{+}_1}{\nu m}\right)\tilde{J}(N),
\en
which is then combined with Eq.~(\ref{K1}) into
\be
\int^1_0dx(1-x)^{N-1}K\otimes J
&=&-\frac{\alpha_sC_F}{\pi}\ln \frac{\mu\bar N}{P^+_1}\tilde{J}(N).
\en
It is seen that the infrared regulator $m^2$ has disappeared in the above combination.

\begin{figure}[!]
\begin{center}
\includegraphics[scale=0.8]{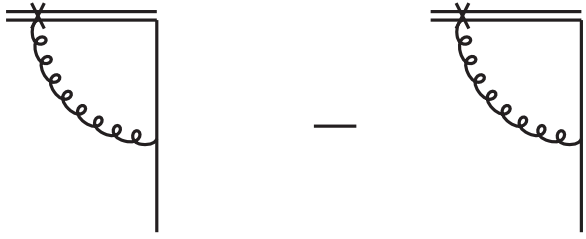}
\vspace{-19.5cm} \caption{$\mathcal{O}(\alpha_s)$ hard function $G$.}\label{fig3}
\end{center}
\end{figure}

The first diagram for the $\mathcal{O}(\alpha_s)$ hard function $G$ in Fig.~\ref{fig3}
contributes
\be
G_1&=&-ig^2C_F\int\frac{d^{4}l}{(2\pi)^{4}}tr\left[\frac{\nsl_+\nsl_-}{4}\gamma_{\nu}
\frac{\psl_2-\ksl+\lsl}{(P_2-k+l)^2}\right]\frac{g^{\mu\nu}}{l^2-m^2}\frac{\hat{n}_\mu}{n\cdot l}\non
&=&-\frac{\alpha_sC_F}{2\pi}\ln\frac{4\nu^2 P^{-2}_2}{m^2 e}.
\en
We have dropped the small momentum $k$, to which the hard function is not sensitive. Instead,
the infrared regulator $m^2$ is introduced, whose dependence will be removed by
the subtraction below. Note that the above expression is free of an ultraviolet divergence.
The final result for $G$ in Fig.~\ref{fig3} is
\be
G=G_1-K_1-\delta G
=-\frac{\alpha_sC_F}{\pi}\ln\frac{2\nu^2 P^-_2}{\sqrt{e}\mu},
\en
with the additive counterterm $\delta G=-\delta K$, where the subtraction
$K_1$ avoids double counting of the soft contribution.

\subsection{Solution to Evolution Equation}

The functions $K$ and $G$ involve only single soft and ultraviolet logarithms, respectively,
so they can be treated by renormalization-group (RG) methods:
\be
\mu\frac{d}{d\mu}K=-\lambda_K=-\mu\frac{d}{d\mu}G,
\en
where the anomalous dimension $\lambda_K=\mu d\delta K/d\mu$
is given up to two loops by \cite{Kodaira:1981nh}
\be
\lambda_K=\frac{\alpha_s}{\pi}C_F+(\frac{\alpha_s}{\pi})^2C_F
\left[C_A\left(\frac{67}{36}-\frac{\pi^2}{12}\right)-\frac{5}{18}n_f\right],
\en
with $C_A=3$ being a color factor and $n_f$ the number of quark flavors. We allow the scale $\mu$ to evolve
to $P_1^+/\bar N$ in $K$ and to $2\nu^2 P^-_2/\sqrt{e}$ in $G$, obtaining the RG solution
of $K+G$,
\be
&&K\left(\frac{P^+_1}{\mu \bar N},\alpha_s(\mu)\right)
+G\left(\frac{2\nu^2 P^-_2}{\sqrt{e}\mu},\alpha_s(\mu)\right)\non
&=&K(1,\alpha_s(P^+_1/\bar N))+G(1,\alpha_s(2\nu^2 P^-_2/\sqrt{e}))
-\int^{2\nu^2 P^-_2/\sqrt{e}}_{P^+_1/\bar N}\frac{d\mu}{\mu}\lambda_K(\alpha_s(\mu))\non
&=&-\int^{2\nu^2 P^-_2/\sqrt{e}}_{P^+_1/\bar N}\frac{d\mu}{\mu}\lambda_K(\alpha_s(\mu)).
\en

Substituting the above evolution kernel into Eq.~(\ref{evo}), we solve for the jet function
\be
\tilde{J}(N)&=&\tilde{J}_{in}(N)\exp\left[-
\int^{\nu^2}_{\sqrt{e}P_1^+/(2\bar N P_2^-)}\frac{d\bar\nu^2}{2\bar\nu^2}
\int^{2\bar\nu^2 P^-_2/\sqrt{e}}_{P^+_1/\bar N}\frac{d\mu}{\mu}\lambda_K(\alpha_s(\mu))\right]\non
&=&\tilde{J}_{in}(N)\exp\left[-\frac{1}{2}\int^{2\nu^2 P_2^-/(\sqrt{e}P_1^+)}_{1/\bar N}\frac{dy}{y}
\int^{y}_{1/\bar N}\frac{d w}{w}\lambda_K(\alpha_s(wP_1^+))\right],\label{th1}
\en
with the variable changes $\mu=w P_1^+$ and $\bar\nu^2=\sqrt{e}P_1^+ y/(2P_2^-)$.
The initial condition $\tilde{J}_{in}(N)$ for the jet function is determined via matching:
we expand Eq.~(\ref{th1}) to ${\cal O}(\alpha_s)$ for a fixed coupling constant,
and compare it with Eq.~(\ref{j2}) to get
\be
\tilde{J}_{in}(N)=\frac{1}{N}\left[1-\frac{\alpha_sC_F}{4\pi}(\frac{3\pi^2}{2}-\frac{1}{4})\right].
\en
If an order-unity constant $C$ is introduced into the exponent,
the initial condition will be modified accordingly:
\be
\tilde{J}(N)&=&\frac{1}{N}\left[1-\frac{\alpha_sC_F}{4\pi}(\frac{3\pi^2}{2}-\frac{1}{4}+\ln^2 C)\right]\non
& &\times\exp\left[-\frac{1}{2}\int^{2\nu^2 P_2^-/(\sqrt{e}P_1^+)}_{C/\bar N}\frac{dy}{y}
\int^{y}_{1/\bar N}\frac{d w}{w}\lambda_K(\alpha_s(wP_1^+))\right].
\en

For a heavy-to-light transition at maximal recoil, we have $P_1^+=P_2^-=m_B/\sqrt{2}$,
$m_B$ being the $B$ meson mass.
Choosing the factor $\nu^2=1/2$, ie., $\xi^2=Q^2$, and
neglecting the running of the coupling constant, we derive the jet function in the Mellin space
\be
\tilde J(\bar N)=\tilde{J}_{in}(N)
\exp\left[-\frac{\lambda_K}{4}\left(\ln^2\bar N-\ln\bar N+\frac{1}{4}\right)\right].
\en
This is the improvement of the threshold resummation with a fixed coupling constant
to the NLL accuracy.
The inverse Mellin transformation brings the jet function back to the momentum fraction space,
\be
J(x)&=&\int^{c+i\infty}_{c-i\infty}\frac{dN}{2\pi i}(1-x)^{-N}\tilde J(N)\non
&=&J_0\int^{\infty}_{-\infty}\frac{dt}{\pi}(1-x)^{\exp(t-\gamma_E+1/2)}
\sin(\frac{\lambda_K \pi t}{2})\exp(-\frac{\lambda_Kt^2}{4}),\label{fix}
\en
with the coefficient
\be
J_0=-\left[1-\frac{\alpha_sC_F}{4\pi}(\frac{3\pi^2}{2}-\frac{1}{4})\right]\exp(\frac{\lambda_K\pi^2}{4}).
\en
In the above formula $c$ is an arbitrary real constant larger than the real
parts of all the poles of the integrand, the variable change
$N=\exp(t+i\pi)$ ($N=\exp(t-i\pi)$) has been adopted for the piece of
contour above (below) the branch cut in Fig.~3 of \cite{Li:2001ay}, and the further variable change
$t+\gamma_E-1/2\to t$ has been made.
It is found that Eq.~(\ref{fix}) exhibits the features similar to those of the LL jet fucntion \cite{Li:2001ay}:
it vanishes as $x\to 0$ and $x\to 1$, and it is normalized to unity up to corrections of $\mathcal{O}(\alpha_s)$.

Next we take into account the running effect of the coupling constant by inserting
$\alpha_s(\mu)=4\pi/[\beta_0\ln (\mu^2/\Lambda^2)]$
into Eq.~(\ref{th1}), with $\beta_0=11-2n_f/3$ and the QCD scale $\Lambda\equiv\Lambda_{\rm QCD}$,
and arrive at
\be
\tilde J(N)&=&\tilde{J}_{in}(N)\exp\left[-\frac{C_F}{\beta_0}\left(
\ln\frac{P_2^{-}}{\sqrt{e}\Lambda}\ln\ln\frac{P_2^{-}}{\sqrt{e}\Lambda}
-\ln\frac{P_2^{-}}{\sqrt{e}P_1^{+}}\right)\right]\non
& & \times \exp\left[\frac{C_F}{\beta_0}\left(
\ln\frac{P_2^{-}}{\sqrt{e}\Lambda}\ln\ln\frac{P_1^{+}}{\Lambda\bar N}
+\ln\bar N\right)\right],\label{fin}
\en
for $\nu^2=1/2$. Note that the above expression vanishes as $\bar N$
approaches to the Landau pole, $\bar N\to P_1^+/\Lambda$, namely, as
$N \approx 8.4$ for $m_B=5.28$ GeV and $\Lambda=0.25$ GeV \cite{Li:2005kt}.

\section{NUMERICAL RESULTS}

In this section we examine the NLL threshold resummation effect on
various CP asymmetries in the $B\to K\pi$ decays. The first step is to convert the jet function
in the Mellin space to the momentum fraction space, which is usually done through the
inverse Mellin transformation defined by the first line of Eq.~(\ref{fix}).
Due to the existence of the Landau pole,
an extrapolation of Eq.~(\ref{fin}) in the large $N$ region is necessary for avoiding this
singularity, which then introduces theoretical uncertainty. On the other hand, it has been
observed \cite{LL2} that the threshold resummation effect is mainly governed by the behavior
of Eq.~(\ref{fin}) at intermediate $N$ for currently accessible energy scales.
Therefore, we will employ the best fit method proposed in \cite{TLS}, instead of the inverse
Mellin transformation, for the aforementioned conversion: the Mellin
transformation of a parametrized jet function is fit to Eq.~(\ref{fin}) in the intermediate
$N$ region.

\begin{figure}[!]
\begin{center}
\vspace{-10cm}
\hspace{76cm}
{\centering
\includegraphics[width=0.8\textwidth,natwidth=290,natheight=382]{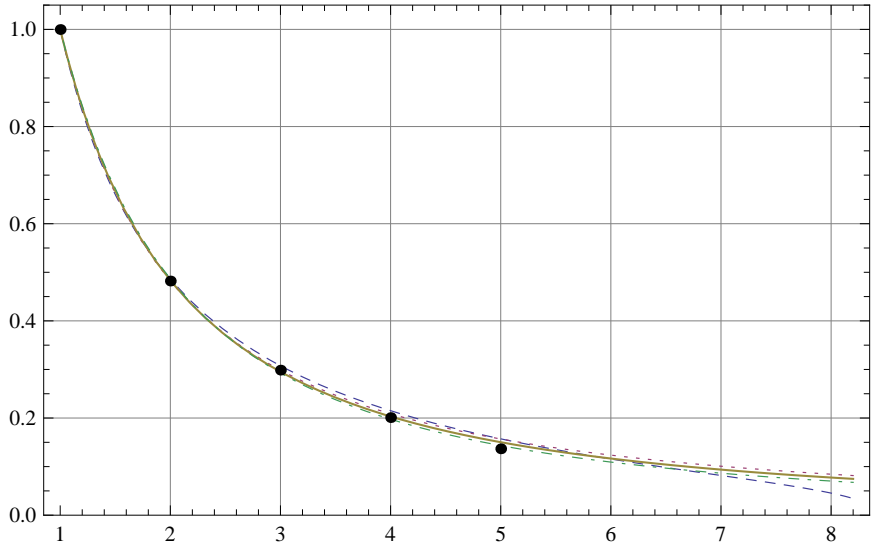}}
 \caption{Jet functions fitted to Eq.~(\ref{fin}) with
LO $\tilde{J}_{in}$ in the ranges $N=1$-3 (dotted line), $N=1$-4
(solid line), and $N=1$-5 (dash-dotted line), where the horizontal
(vertical) axis is labelled by $N$ ($\tilde{J}(N)$). The exact $N$
dependence of Eq.~(\ref{fin}) (dashed line) is also displayed for
comparison.}\label{ful}
\end{center}
\end{figure}

We parametrize the jet function in the momentum fraction space as
\begin{eqnarray}
J(x)=\frac{\Gamma(\alpha+\beta+2)}{\Gamma(\alpha+1)\Gamma(\beta+1)}x^\alpha (1-x)^\beta,\label{par}
\end{eqnarray}
which is motivated by the feature of Eq.~(\ref{fix}), ie., vanishing as $x\to 0$ and $x\to 1$.
The prefactor has been introduced to obey the normalization $\int dx J(x)=\tilde J(1)=1$.
It implies that we have chosen the initial condition at LO, $\tilde{J}_{in}(N)=1/N$, since
we intend to focus on effects from the resummaton. To be consistent, the $B\to K\pi$
factorization formulas with the LO hard kernels will be
adopted for the numerical study below. We mention that Eq.~(\ref{fin}) is roughly, but
not exactly, equal to unity as $N=1$ even with the LO $\tilde{J}_{in}(N)$. The equality can be made
exact by choosing $\nu^2=\exp(1/2-\gamma_E)/2\approx 0.47$, quite close to $\nu^2=1/2$ taken
in this work. The Mellin transformation of Eq.~(\ref{par}) is then fit to
Eq.~(\ref{fin}) in the intermediate $N$ region, and its deviation from Eq.~(\ref{fin}) at
large $N$ is regarded as an extrapolation to avoid the Landau singularity.

The best fits to Eq.~(\ref{fin}) for $n_f=4$ in the ranges from $N=1$ to 3, $N=1$ to 4,
and $N=1$ to 5 produce the curves displayed in Fig.~\ref{ful}, which exhibit
good agreement with Eq.~(\ref{fin}) at intermediate $N$, and start to deviate from Eq.~(\ref{fin})
as $N>6$. We take the jet function from the $N=1$-4 fit with the parameters $\alpha=0.58$ and $\beta=0.47$
to generate our results, and those from the $N=1$-3 fit
($\alpha=0.43$ and $\beta=0.33$) and from the $N=1$-5 fit ($\alpha=0.76$ and $\beta=0.67$)
to estimate the theoretical uncertainty. The similarity among the three fitted jet functions
guarantees that the uncertainty from avoiding the Landau singularity is not severe
in our best fit method. Compared to the LL jet function
$J(x)\propto [x(1-x)]^{0.3}$ \cite{TLS}, the NLL one provides stronger suppression
at the end points of $x$, with which particles involved in the hard decay kernels tend to be
more off-shell, and the perturbative analysis of the $B\to K\pi$ decays is expected
to be more reliable.

\begin{table}
\caption{Direct and mixing-induced CP asymmetries in the $B\to K\pi$ decays. }
\begin{center}
\begin{tabular}{c|c|c|c|c|c}
\hline\hline &Data & LL& NLL ($N=1$-3) & NLL ($N=1$-4) & NLL ($N=1$-5)\\
\hline
$A_{CP}(K^{0}\pi^-)$& $-0.017\pm0.016$ &$-0.012$ &$-0.012$ &$-0.015$ &$-0.019$\\
$A_{CP}(K^{-}\pi^0)$& $0.037\pm0.021$  &$-0.085$ &$-0.063$ &$-0.070$  &$-0.076$\\
$A_{CP}(K^{-}\pi^+)$& $-0.082\pm0.006$  &$-0.12$ &$-0.090$ &$-0.097$  &$-0.10$\\
$A_{CP}(K^{0}\pi^0)$& $0.00\pm0.13$ &$-0.024$    &$-0.019$ &$-0.018$  &$-0.018$\\
$S_{K^{0}\pi^0}$    &$0.58\pm0.17$  &$0.683$     &$0.688$  &$0.704$   &$0.720$\\
\hline\hline
\end{tabular}\label{tab2}
\end{center}
\end{table}

For the Cabibbo-Kobayashi-Maskawa matrix elements, we take the Wolfenstein parametrization with
the values $A=0.836\pm0.015$, $\lambda=0.22453\pm0.00044$, $\bar\rho=0.122^{+0.018}_{-0.017}$
and $\bar\eta=0.355^{+0.012}_{-0.011}$ \cite{Tanabashi:2018oca}. The hadronic inputs,
including meson masses and decay constants, meson distribution amplitudes, chiral scales $m_0$
associated with the pion and kaon twist-3 distribution amplitudes, and the QCD scale $\Lambda_{\rm QCD}$ are
the same as in \cite{Li:2005kt}. The factorization formulas for the relevant
$B\to K\pi$ decay amplitudes with the LO hard kernels are also referred to \cite{Li:2005kt}.
The outcomes for the CP asymmetries under the LL and NLL threshold resummations
are listed in Table~\ref{tab2}, in which the values
in the column labelled by LL well reproduce the corresponding ones in \cite{Li:2005kt}.
It is found that the NLL effect enhances the direct CP asymmetry $A_{CP}(K^{0}\pi^-)$ by 25\%,
and decreases the other three direct CP asymmetries by 20-25\% relative to the LL results.
The mixing-induced CP asymmetry $S_{K^{0}\pi^0}$, increasing by only 3\%, is less sensitive to
the replacement of the jet function. It is understandable, because this observable
is supposed to be close to $\sin(2\phi_1)$, $\phi_1$ being the weak phase, in penguin-dominated
modes like $B\to K\pi$.
The comparison of the column labelled by NLL ($N=1$-4) with those
labelled by NLL ($N=1$-3) and NLL ($N=1$-5) indicates that the theoretical uncertainty is under control:
except $A_{CP}(K^{0}\pi^-)$, whose uncertainty amounts up to 20\%, the other CP
asymmetries change by lower than 10\%.


\section{CONCLUSION}

In this paper we have improved the LL threshold resummation for exclusive $B$ meson decays
to the NLL accuracy. The recipe contains the computation of the
one-loop jet function factorized out of decay amplitudes,
the derivation of the evolution kernels, the matching of the resummation
formula to the one-loop jet function, and the inclusion of the running effect of the
coupling constant. It has been observed that the NLL threshold resummation suppresses
the end-point region with $x\sim 0$ more strongly than the LL one.
Since we focused on the resummaton effect, we did not take into account the NLO piece
in the initial condition of the jet function. For consistency, we worked on the PQCD
factorization formulas for the $B\to K\pi$ decays with the LO hard kernels.
It has been explained that the different LL and NLL threshold resummation
effects can be compared unambiguously through the investigation of the CP asymmetries.
We have shown that the replacement of the LL jet function by the
NLL one causes about 20-25\% variation of the $B\to K\pi$ direct CP asymmetries,
which is not negligible for precision analyses for $B$ meson decays. On the
contrary, the mixing-induced CP asymmetry almost remains untouched under the
above replacement. Moreover, the theoretical uncertainty from the inverse Mellin
transformation of the threshold resummation is under control.

The implementation of the NLL threshold resummation derived here in the PQCD approach to
exclusive $B$ meson decays is nontrivial, and demands more efforts. As pointed out in
the Introduction, the threshold resummation modifies hard decay
kernels by including partial higher order contributions, so hadron distribution amplitudes should be
adjusted accordingly. In principle, it is more appropriate to execute this task
in a global study of many two-body hadronic $B$ meson decay modes.
A global fit to available data based on the PQCD approach
with the NLL threshold resummation will be attempted in near
future.

{\bf Acknowledgement}

We thank S. Mishima for useful discussions.
This work was supported in part by MOST of R.O.C. under Grant No.
MOST-107-2119-M-001-035-MY3, and by NSFC under Grant No.11347030.



\begin{thebibliography}{99}
\bibitem{SHB} A. Szczepaniak, E.M. Henley, and S.J. Brodsky,
Phys. Lett. B {\bf 243}, 287 (1990).
\bibitem{ASY} R. Ahkoury, G. Sterman, and Y.P. Yao, Phys. Rev. D
{\bf 50}, 358 (1994).
\bibitem{BF} M. Beneke and T. Feldmann, Nucl. Phys. {\bf B592}, 3
(2000).

\bibitem{BBNS}
  M. Beneke, G. Buchalla, M. Neubert, and C.T. Sachrajda,
  Phys. Rev. Lett. {\bf 83}, 1914 (1999);
  Nucl. Phys. {\bf B591}, 313 (2000); {\it ibid}. {\bf B606}, 245 (2001).

\bibitem{Manohar:2006nz}
  A.~V.~Manohar and I.~W.~Stewart,
  Phys.\ Rev.\ D {\bf 76}, 074002 (2007).

\bibitem{LY1} H.~n. Li and H.~L. Yu, Phys. Rev. Lett. {\bf 74},
4388 (1995); Phys. Rev. D {\bf 53}, 2480 (1996).
\bibitem{KLS} Y.~Y. Keum, H.~n. Li and A.~I. Sanda,
Phys. Lett. B {\bf 504}, 6 (2001); Phys. Rev. D {\bf 63},
054008 (2001); Y.~Y. Keum and H.~n. Li, Phys. Rev. D {\bf 63}, 074006 (2001).
\bibitem{LUY} C.~D. Lu, K. Ukai, and M.~Z. Yang, Phys. Rev. D {\bf 63},
074009 (2001).
\bibitem{TLS}
T.~Kurimoto, H.~n. Li and A.~I. Sanda,
Phys.\ Rev.\ D \textbf{65}, 014007 (2002).

\bibitem{Li:2001ay}
  H.~n.~Li,
  Phys.\ Rev.\ D {\bf 66}, 094010 (2002).

\bibitem{KPY} G.~P. Korchemsky, D. Pirjol, and T.~M. Yan,
Phys. Rev. D {\bf 61}, 114510 (2000).

\bibitem{Li:2002mi}
  H.~n.~Li and K.~Ukai,
  Phys.\ Lett.\ B {\bf 555}, 197 (2003).

\bibitem{Li:2005kt}
H.~n. Li, S.~Mishima and A.~I. Sanda,
Phys.\ Rev.\ D \textbf{72}, 114005 (2005).

\bibitem{Li:1996gi}
H.~n.~Li,
Phys. Rev. D \textbf{55}, 105 (1997).

\bibitem{Li:2013ela}
H.~n.~Li,
Phys. Part. Nucl. \textbf{45}, 756 (2014).

\bibitem{Li:2012md}
H.~n.~Li, Y.~L.~Shen and Y.~M.~Wang,
JHEP \textbf{02}, 008 (2013).


\bibitem{Li:2013xna}
H.~n.~Li, Y.~L.~Shen and Y.~M.~Wang,
JHEP \textbf{01}, 004 (2014).

\bibitem{Kodaira:1981nh}
J.~Kodaira and L.~Trentadue,
Phys.\ Lett.\ B \textbf{112}, 66 (1982).


\bibitem{LL2} H.~L. Lai and H.~n. Li, Phys. Lett. B. {\bf 471}, 220 (1999).




\bibitem{Tanabashi:2018oca}
M.~Tanabashi \textit{et al.} [Particle Data Group],
Phys. Rev. D \textbf{98}, 030001 (2018).




\end{thebibliography}
\end{document}